\documentclass[pre,twocolumn,superscriptaddress]{revtex4-1}
\usepackage{graphicx}
\usepackage{dcolumn}
\usepackage{bm}
\usepackage{amsmath}
\usepackage{amsfonts}
\usepackage{soul}

\begin{document}

\title{Mean-field interactions between living cells in linear and nonlinear elastic matrices}

\author{Chaviva Sirote}
\affiliation{Department of Biomedical Engineering, Tel Aviv University, Tel Aviv 69978, Israel}

\author{Yair Shokef}
\email{shokef@tau.ac.il}
\homepage{https://shokef.tau.ac.il}
\affiliation{School of Mechanical Engineering, Tel Aviv University, Tel Aviv 69978, Israel}
\affiliation{Sackler Center for Computational Molecular and Materials Science, Tel Aviv University, Tel Aviv 69978, Israel}
\affiliation{Center for Physics and Chemistry of Living Systems, Tel Aviv University, Tel Aviv 69978, Israel}

\begin{abstract}
Living cells respond to mechanical changes in the matrix surrounding them by applying contractile forces that are in turn transmitted to distant cells. We calculate the mechanical work that each cell performs in order to deform the matrix, and study how that energy changes when a contracting cell is surrounded by other cells with similar properties and behavior. We consider simple effective geometries for the spatial arrangement of cells, with spherical and with cylindrical symmetries, and model the presence of neighboring cells by imposing zero-displacement at some distance from the cell, which represents the surface of symmetry between neighboring cells. In linear elastic matrices, we analytically study the dependence of the resulting interaction energy on the geometry and on the stiffness and regulatory behavior of the cells. For cells that regulate the active stress that they apply, in spherical geometry, the deformation inside the cell is pure compression thus the interaction depends only on their bulk modulus, while in cylindrical geometries the deformation includes also shear and the interaction depends also on their shear modulus. In nonlinear, strain stiffening matrices, our numerical solutions and analytical approximations show how in the presence of other cells, cell contraction is limited due to the divergence of the shear stress.
\end{abstract}

\maketitle

\section{Introduction}

Living cells embedded in, or adhered on an elastic environment transmit mechanical forces through deformations of their surrounding solid medium~\cite{SchwarzRMP2013}. Thus their active contraction may be felt by non-contacting cells through the propagation of stress and strain fields via this matrix. Such mechanical interactions between cells and their surrounding environments, as well as matrix-mediated cell-cell interactions are important for many biological processes, such as stem-cell differentiation~\cite{Engler2006}, wound healing~\cite{Poujade2007}, embryonic development~\cite{Montrell2003}, cell division~\cite{Lesman2014, Abuhattum2015}, cancer metastasis~\cite{Freidl2009, Gal2012}, and cell-cell biochemical communication~\cite{Gomez2019, Gomez2020, Jung2020}. Recent experiments in synthetic setups or geometries enable to isolate and study these mechanical interactions~\cite{ReinhartKing2008, Winer2009, Buxboim2010, Mohammadi2014, Lesman2015Interface, Lesman2015Integrative}.

Theoretically, cell contraction is customarily modeled by assuming that cells generate active force dipoles~\cite{Schwarz2002}, namely that they apply on the elastic medium pairs of equal-magnitude and oppositely-pointing forces. The interaction between distant such force dipoles may be described in terms of the excess elastic energy stored in the medium in a situation with interacting cells compared to the total energy of these cells without interactions. Interaction energies between active force dipoles have been studied in the past in the context of atoms adsorbed to surfaces~\cite{Lau1977}, and more recently in the cell mechanics context, with studies ranging from simple, linear dipoles comprised of pairs of point forces~\cite{Bischofs2003, Bischofs2004, Bischofs2005, Bischofs2006} to more complex continuous shapes \cite{Golkov2019}.

To simplify the description of actively contractile cells, cells within a three-dimensional medium~\cite{Lesman2014, Lesman2015Integrative} have been modeled as spherical force dipoles~\cite{Shokef2012, Xu2015, BenYaakov2015, Wang2020}. Namely, the mechanical activity of each cell is described by an isotropic distribution of radial external forces that are applied on the surface of a sphere within a three-dimensional elastic medium. This may be thought of as a continuous collection of linear force dipoles distributed isotropically on the surface of this sphere, with each such linear force dipole applying equal and opposite forces at two opposing points on the sphere. This spherical geometry implies spherical symmetry for the elastic fields around a cell, and thus simplifies the partial differential equations of elastic equilibrium that should be solved to a one-dimensional or ordinary differential equation for the radial displacement as a function of the radial coordinate.

Along these lines, interactions between biological cells have been modeled by imposing a zero-displacement boundary condition at some distance from this spherical cell~\cite{BenYaakov2015}. The justification for this is to first assume that the spatial distribution of cells in their elastic environment is on a periodic array. In such case, by symmetry, the normal displacement on the surface of each periodic unit-cell of this array should vanish. A reasonable mean-field approximation would then be to replace the actual polyhedral periodic unit-cell with a spherical unit-cell with zero displacement on its boundary, which results in a simple one-dimensional geometry also for the study of cell-cell interactions. Note that the periodic unit-cell of an array of biological cells and the spherical domain approximating this unit-cell should not be confused with the biological cells.

Cell-cell elastic interactions are similar, but not identical to electrostatic interactions between induced electric dipoles~\cite{Israelachvili1992, Schwarz2002, Bischofs2004, SchwarzRMP2013, Golkov2017}, and the interaction energy scales algebraically with distance. Interestingly, the sign of the interaction energy is set by the homeostatic behavior of the cells; cells regulating the displacement on their boundary repel each other, and cells regulating the force that they apply are mechanically attracted to each other~\cite{BenYaakov2015}. Elaborate solutions, which include the interference between the angular dependence of the deformation fields around two spheres recover these mean-field results, both in terms of the sign and of the exponents of the power-law decay in the magnitude of the interaction energy~\cite{Golkov2017}.

In this Paper we use the aforementioned mean-field approach to study additional geometries, namely not only spherical symmetry, but also two simple cylindrical setups~\cite{Sopher2018} that relate to cells on substrates or to elongated cells in a three-dimensional matrix. In doing that, we provide further insight on matrix-mediated elastic interactions in more complex geometries, for which analytical results are not known. Furthermore, we include in our calculations the stiffness of the interior the cell, and not only that of the surrounding matrix. Finally, and most importantly, we extend our analysis to cells in nonlinear, strain-stiffening media~\cite{Shokef2012}.

\section{Model and Interaction Energy} \label{Sec:Model}

\begin{figure}[h]
\includegraphics[clip , trim=1cm 0cm 9cm 0cm , width=\columnwidth]{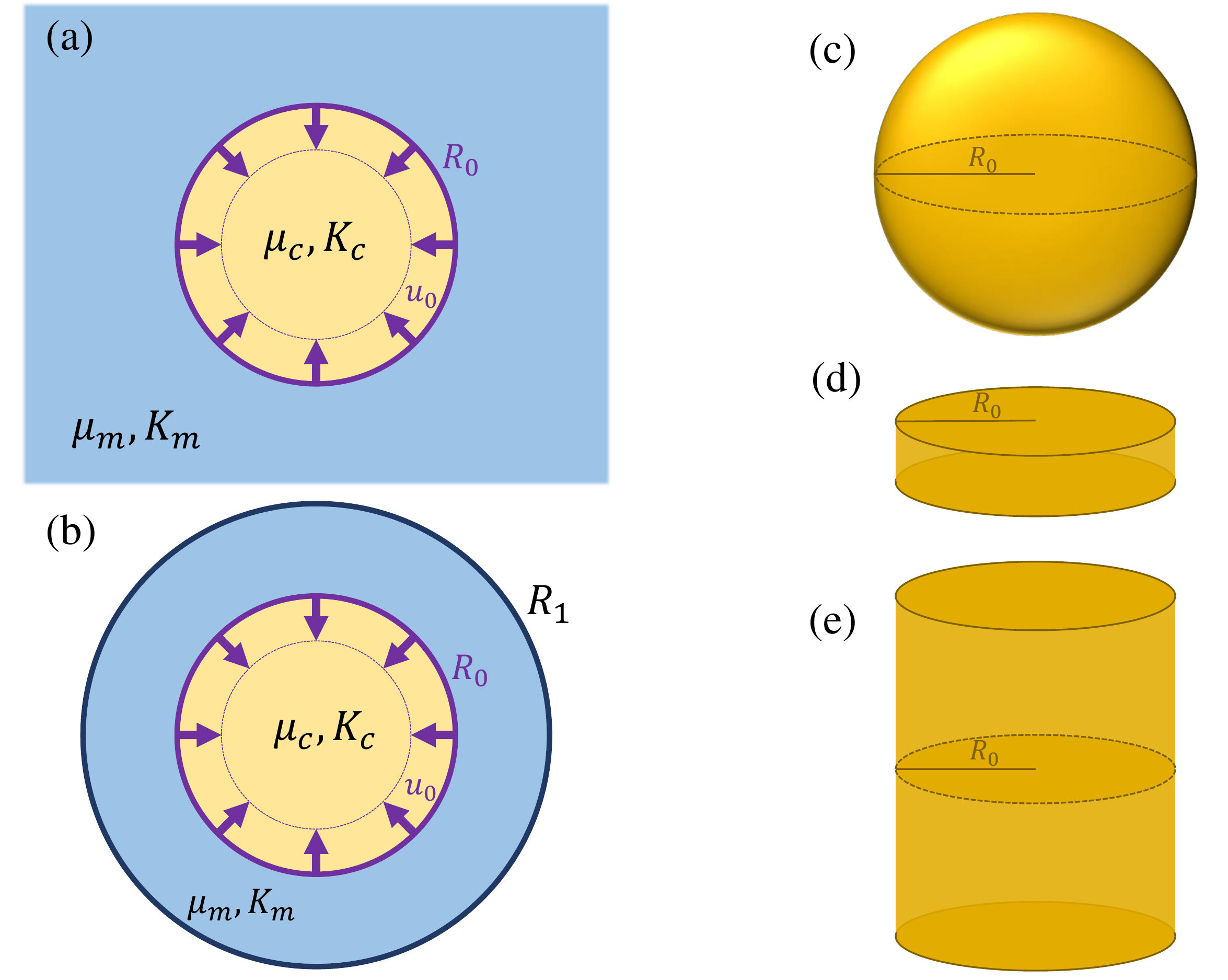}
\caption{Schematic of a single cell in an infinite matrix (a) and in a bounded region (b). The cell may be a sphere (c), a flat cylinder (d) for plane stress, or a long cylinder (e) for plane strain.}
\label{Fig:circledrawing}
\end{figure}

We analyze two geometries, one in which the contracting cell is assumed to be a sphere of radius $R_0$, and the other in which it is assumed to be a cylinder of radius $R_0$. In both geometries, we model the presence of neighboring cells by imposing a rigid wall with zero normal displacement at some larger radius $R_1$, see Fig.~\ref{Fig:circledrawing}. This radius represents half of the distance to nearby cells, since in realistic geometries, by symmetry the normal displacement should vanish at that position. In our model, the biological cell has linear elastic response characterized by shear modulus $\mu_c$ and bulk modulus $K_c$. In Sec.~\ref{Sec:linear} we will assume that the matrix is also linearly elastic with shear modulus $\mu_m$ and bulk modulus $K_m$, while in Sec.~\ref{Sec:nonlinear} we will include nonlinear strain-stiffening of the matrix.

The elastic energy stored in the cell and in the matrix together is equal to the mechanical work performed by the active, or external forces, which act only at the surface of the cell. Here, these are radial forces that act at the radius $R_0$ and generate a radial displacement $u_0$ there. We assume that in the absence of these active forces, the elastic medium is relaxed, and hence we calculate the work done on the matrix by integrating over the adiabatic process of building up the displacement $u_0$, starting from this relaxed state~\cite{Golkov2019}
\begin{equation}
E = \int_0^{u_0} g \tau(w) dw . \label{Eq:work-def}
\end{equation}
Here $w$ denotes the displacement at $R_0$ during this process, and $\tau(w) = \sigma_c(R_0) - \sigma_m(R_0)$ is the active stress applied by the cell, namely the discontinuity in the radial component of the stress tensor on the surface of the cell when the displacement there is equal to $w$. We use $\sigma_c$ and $\sigma_m$ to denote the radial component $\sigma_{rr}$ of the stress tensor in the cell and in the matrix, respectively. The geometrical factor relating stress to force is equal to $g = 4 \pi R_0^2$ in the spherical geometry. In the cylindrical geometry we use $g = 2 \pi R_0$ to obtain the force per unit length of the cylinder, and subsequently the energy per unit length. Since much of our analysis and discussion below is performed simultaneously for the two geometries, for brevity we will use the term energy and the symbol E, for energy in the spherical geometry and for energy per unit length in the cylindrical geometry. If the medium is linearly elastic (Sec.~\ref{Sec:linear}), there is a linear proportionality between displacement and active stress, $\tau(w) \propto w$, and Eq.~(\ref{Eq:work-def}) reduces to $E=\frac{1}{2}g\sigma_0 u_0$, where $\sigma_0=\tau(u_0)$ is the active stress, or stress difference on the surface of the cell when the displacement there is equal to $u_0$~\cite{BenYaakov2015, Golkov2017}. However, for nonlinear media (Sec.~\ref{Sec:nonlinear}), the entire dependence of $\tau$ on $w$ up to the actual displacement $u_0$ is required in order to perform the integration in Eq.~(\ref{Eq:work-def}).

We define the interaction energy as the added work that the cell has to perform due to the presence of neighboring cells. Within our mean-field approximation, this is equal to the difference 
\begin{equation}
\Delta E = E_1 - E_\infty \label{eq: interactionenergy}
\end{equation}
between the elastic energy $E_1$ stored in the medium when there is a zero-displacement boundary condition at $R_1$ (Fig.~\ref{Fig:circledrawing}b) and the elastic energy $E_\infty$ of the same cell in an unbounded matrix (Fig.~\ref{Fig:circledrawing}a). Care should be given to the meaning of this placement of the same cell in two different mechanical environments. Obviously when identical cells have different environments, they behave differently~\cite{Engler2006, Brown1998, Freyman2002, Saez2005, Ghibaudo2008, De2008, He2014}. In our case, the confining geometry alters the relation between the active stress $\tau$ and the displacement $u_0$. Thus cells in the confined and in the unbounded geometries cannot simultaneously have both the same active stress and the same deformation. We will consider two extreme scenarios: i) displacement regulation, for which cells are assumed to be biologically programmed to generate a given displacement $u_0$, regardless of what active stress they need to apply in order to reach that deformation, and ii) stress regulation, for which cells apply a given active stress $\sigma_0$, regardless of the displacement they manage to generate.

For cells regulating the displacement on their surface, the far-field zero-displacement boundary condition imposed by the presence of neighboring cells causes cells to apply a higher active stress in order to generate a given contraction. Thus we expect the interaction energy to be positive, representing repulsion between cells. For stress regulation, on the other hand, a smaller displacement will be generated in the presence of other cells, thus leading to a negative interaction energy, or to attraction~\cite{BenYaakov2015, Golkov2017}. We consider cells that are embedded in a solid surrounding, thus they cannot easily move due to these repulsive or attractive interactions. However, we suggest that these interactions are the mechanical cause for instance for cells to send protrusions one toward each other or rather away from each other~\cite{Mohammadi2014} or for cells to generate bands of densification and alignment of the network of fibers around them~\cite{Gomez2019, Natan2021}, and also for migratory behavior of cells on substrates~\cite{ReinhartKing2008}. Alternatively, our results could be used to explain when do cells regulate their displacement and when do they regulate their active stress, depending on the energetic cost for doing so.

Our strategy for calculating the interaction energy for the different cases described above is as follows: We first assume a displacement $u_0$ on the boundary of the cell, and solve for the displacement field. Using the resulting strain, we calculate the stress on the boundary of the cell, and from that the active stress $\sigma_0$ that the cell has to apply to generate the displacement $u_0$. We then use these to calculate the elastic energy stored in the deformation. For displacement regulation we subtract the resulting energies in the bounded and in the unbounded geometries to get the interaction energy, Eq.~(\ref{eq: interactionenergy}) at a given displacement $u_0$. For stress regulation, we combine our results for $E$ vs. $u_0$ and for $\sigma_0$ vs. $u_0$ to obtain $E$ vs. $\sigma_0$, and then use Eq.~(\ref{eq: interactionenergy}) to get the interaction energy at a given active stress $\sigma_0$. This procedure has been applied previously~\cite{BenYaakov2015} for an empty spherical cell in a linearly elastic medium, with the focus on the asymptotic behavior when cells are far apart from each other, which translates to $R_1 \gg R_0$ in our model. Here, we provide the full dependence on distance, and also: i) include a passive, elastic rigidity of the interior of the cell, ii) solve for additional cases with cylindrical symmetry, and iii) study the effects of nonlinear elastic response of the matrix.

\section{Linear Medium} \label{Sec:linear}

Our spherical and cylindrical geometries both assume radial symmetry, such that the displacement vector $\vec{u}(\vec{r}) = u(r) \hat{r}$ is given only by the radial component $u(r)$, which depends only on the radial coordinate $r$. In this section we assume that the mechanical response of both the cell and the surrounding matrix is linearly elastic. This leads to linear ordinary differential equations for $u(r)$, which we analytically solve. Thus we obtain exact analytical expressions for the interaction energy in the different geometries. More complicated nonlinear material models may be considered for the cell and for the matrix, for instance the nonlinear elastic response of the matrix as described in Sec.~\ref{Sec:nonlinear} below. Such models result in nonlinear differential equations that need to be solved in order to obtain the displacement field. However, the spherical and cylindrical symmetries of our geometrical setups reduces the general three-dimensional partial differential equations of nonlinear elasticity to one-dimensional ordinary differential equations, which, if needed, may be solved numerically much more easily.

\subsection{Spherical Cell}

For spherical symmetry (Fig.~\ref{Fig:circledrawing}c), mechanical equilibrium implies~\cite{Bower2009}
\begin{equation}
\frac{d^2u}{dr^2} + \frac{2}{r}\frac{du}{dr} - \frac{2u}{r^2} = 0 \label{eq:ode sph} .
\end{equation}
The general solution to this equation is:
\begin{equation}
u(r)=C r+\frac{D}{r^2}\label{eq:genral shp} ,
\end{equation}
where the constants $C$ and $D$ are set by the boundary conditions. For spherical symmetry, the radial stress is given by~\cite{Bower2009}
\begin{equation}
\sigma_{rr} = \frac{4}{3}\mu \left(\frac{du}{dr}-\frac{u}{r}\right) + K \left(\frac{du}{dr}+\frac{2u}{r}\right) \label{eq:stress sph} ,
\end{equation}
which for the general solution~(\ref{eq:genral shp}) may be written as
\begin{equation}
\sigma_{rr} = 3 K C - \frac{4 \mu D}{r^3} \label{eq:stressFXuo} .
\end{equation}
Thus for spherical symmetry, the $u \propto r$ term in~(\ref{eq:genral shp}) represents pure compression, while the $u \propto 1/r^2$ term represents pure shear.

Inside the cell ($0<r<R_0$) the boundary conditions are $u(0)=0$ and $u(R_0)=u_0$, thus the displacement is given by
\begin{equation}
u_c = \frac{u_0 r}{R_0} , \label{eq:dispFI}
\end{equation}
which contains only compression, and the stress in the cell is uniform and equal to
\begin{equation}
\sigma_{c}=3 K_{c} \frac{u_0}{R_0} \label{eq:15} ,
\end{equation}
which depends only on the bulk modulus of the cell and not on its shear modulus. 

In the surrounding matrix ($r>R_0$), the displacement at the cell border is $u(R_0)=u_0$. For the unbounded case, the second boundary condition is that $u=0$ at $r=\infty$, leading to
\begin{equation}
u_{m,\infty} = \frac{u_0 R_0^2}{r^2} \label{eq:ShpISoEmp} ,
\end{equation}
which contains only shear. The stress in the matrix in this case is thus
\begin{equation} 
\sigma_{m,\infty} = - \frac{4 \mu_m u_0 R_0^2}{r^3} = -4 \mu_m \frac{u_0}{R_0} , \label{eq:sigma_m_inf} 
\end{equation}
where the second equality is obtained by setting the position to be on the surface of the cell, $r=R_0$. The total work that an isolated cell performs in order to deform both itself and the matrix that surrounds it is thus
\begin{equation}
E_\infty = 2 \pi u_0^2 R_0 (3K_c+4\mu_m) . \label{eq:E0}
\end{equation}

For the bounded case, the second boundary condition is now $u(R_1)=0$, and the displacement in the matrix is
\begin{equation}
u_{m,1} = \frac{u_0 R_0^2}{R_1^3-R_0^3} \left( \frac{R_1^3}{r^2} - r \right)\label{eq:SSE} ,
\end{equation}
which contains both shear and compression. In this case, the stress in the matrix is
\begin{equation}
\sigma_{m,1} = - \frac{4\mu_m R_1^3 + 3 K_m R_0^3}{R_1^3-R_0^3} \frac{u_0}{R_0} \label{eq:step} , 
\end{equation}
and the energy is
\begin{equation}
E_1 = 2 \pi u_0^2 R_0 \left[ 3K_c + \frac{4\mu_m R_1^3+3 K_m R_0^3}{R_1^3-R_0^3} \right] \label{eq:ttt} .
\end{equation}

\subsubsection{Displacement Regulation}

If the cell generates the same displacement $u_0$ on its boundary regardless of its mechanical environment, the interaction energy is just the difference between the energy~(\ref{eq:ttt}) when it is bounded and the energy~(\ref{eq:E0}) when it is unbounded, namely
\begin{equation}
\Delta E = E_1 - E_\infty = 2 \pi u_0^2 R_0^4 \frac{4\mu_m + 3 K_m}{R_1^3-R_0^3}\label{eq:eee} .
\end{equation}
This result depends on both the bulk and shear moduli of the matrix because both compression and shear are generated in the matrix. It does not, however, depend on the elastic moduli of the cell since for displacement regulation, the cell deforms in the exact same way in the bounded and in the unbounded geometries, thus the same energy is stored in the cell for both geometries. As predicted above, for displacement regulation, the interaction energy is positive, representing repulsion between cell. For large separation between cell ($R_1 \gg R_0$), the interaction energy decays as $1/R_1^3$, which is consistent with a mapping to interactions between induced electric dipoles~\cite{BenYaakov2015, Golkov2017}.

\subsubsection{Stress Regulation}

If the cell generates the same active stress $\sigma_0$ both when it is bounded and when it is unbounded, we need to first express the energies $E_1$ and $E_\infty$ in terms of $\sigma_0$ rather than $u_0$, and only then subtract them to get the interaction energy. Using Eqs.~(\ref{eq:sigma_m_inf}) and (\ref{eq:15}), for an unbounded cell, the displacement on its boundary is given by
\begin{equation}
u_{0,\infty}=\frac{\sigma_0 R_0}{3K_c+4\mu_m} \label{u0_st_ri} ,
\end{equation}
And from Eq.~(\ref{eq:E0}), the energy in this case is given by
\begin{equation}
E_\infty = \frac{2\pi\sigma_0^2 R_0^3}{3K_c+4\mu_m} \label{eq:E0_st_ri} .
\end{equation}
Similarly, from Eqs.~(\ref{eq:step}) and (\ref{eq:15}), the displacement on the boundary of a bounded cell is
\begin{equation}
u_{0,1}=\frac{\sigma_0 R_0 (R_1^3-R_0^3)}{4\mu_mR_1^3  + 3K_mR_0^3 + 3K_c\left(R_1^3-R_0^3\right)} , \label{eq:u1_sr_re}
\end{equation}
and the energy~(\ref{eq:ttt}) in this case is
\begin{equation}
E_1 = \frac{ 2\pi\sigma_0^2 R_0^3 (R_1^3-R_0^3)}{4\mu_mR_1^3  + 3K_mR_0^3 + 3K_c\left(R_1^3-R_0^3\right)} . \label{eq:er1_sr_ri}
\end{equation}
By subtracting (\ref{eq:E0_st_ri}) from (\ref{eq:er1_sr_ri}) we get that for stress regulation the interaction energy is
\begin{equation}
\Delta E = - \frac{2 \pi \sigma_0^2 R_0^6 (4\mu_m+3K_m)}{\left( 4\mu_m + 3K_c \right) \left[ 4\mu_mR_1^3  + 3K_mR_0^3 + 3K_c\left(R_1^3-R_0^3\right)\right]} \label{eq:destressridis} .
\end{equation}
As discussed above, for stress regulation, the interaction energy is negative, representing attraction between cells. Moreover, here the contraction of the cell differs between the bounded and unbounded geometries, therefore the interaction energy depends on the stiffness of the cell. However, for spherical symmetry the deformation of the cell includes only compression, thus the interaction energy depends on the bulk modulus of the cell but not on its shear modulus. The $1/R_1^3$ decay in the magnitude of the interaction energy for large distances between cells is similar to that found above for displacement regulation.

\subsection{Cylindrical Cell}

For cylindrical symmetry, the boundary conditions at the ends of the cylinder affect the stress inside it. If the cylinder is prevented from deforming in the axial direction, the axial strain vanishes $\epsilon_{zz}=0$, this is termed plane strain, and the radial stress reads~\cite{Bower2009}
\begin{equation}
\sigma_{rr}^A = \frac{2}{3}\mu \left(2\frac{du}{dr}-\frac{u}{r} \right) + K \left(\frac{du}{dr}+\frac{u}{r}\right) \label{eq:plane_strain} .
\end{equation}
In order to maintain this state, external axial forces should be applied at the ends of the cylinder. Note that there is no displacement there, hence these forces do not perform any work, and thus do not contribute to the energy. If such forces are not applied, the axial stress is zero $\sigma_{zz}=0$, this is termed plane stress, and the radial stress is given by~\cite{Bower2009} 
\begin{equation}
\sigma_{rr}^E = \frac{2}{3}\mu \left(2\frac{du}{dr}-\frac{u}{r}\right) + \frac{1-4\nu^2}{1-\nu^2} K \left(\frac{du}{dr}+\frac{u}{r}\right) \label{eq:plane_stress} ,
\end{equation}
with the Poisson ratio given by
\begin{equation}
\nu = \frac{3K-2\mu}{2(3K+\mu)} . \label{eq:Poisson}
\end{equation} 
Also for plane stress no work is done due to the axial deformation since no forces are applied in that direction. The simple geometries considered here are not meant to exactly describe the shapes of specific biological cells and their environments. However, we suggest that plane stress (Fig.~\ref{Fig:circledrawing}d) may be more relevant for thin cells in a monolayer or for cells on a substrate, while plane strain (Fig.~\ref{Fig:circledrawing}e) may be more relevant for elongated cells such as neurons, in a three dimensional environment.

Regardless of the aforementioned boundary conditions in the axial direction, cylindrical symmetry implies that the radial displacement satisfies the following differential equation, which ensures mechanical equilibrium~\cite{Bower2009}
\begin{equation}
\frac{d^2u}{dr^2}+\frac{1}{r}\frac{du}{dr}-\frac{u}{r^2} = 0 \label{eq:ode sph-1} .
\end{equation}
The general solution to this equation reads:
\begin{equation}
u(r)=Cr+\frac{D}{r} \label{eq:genral shp-1} ,
\end{equation}
where the constants $C$ and $D$ are set by the boundary conditions in the radial direction. We now generalize the two cases for the axial boundary conditions, plane strain (\ref{eq:plane_strain}) and plane stress (\ref{eq:plane_stress}) to write the radial stress for both cases as
\begin{equation}
\sigma_{rr} = \frac{2}{3}\mu \left(2\frac{du}{dr}-\frac{u}{r} \right) + \hat{K} \left(\frac{du}{dr}+\frac{u}{r}\right) \label{eq:strees cy} ,
\end{equation}
with an effective bulk modulus given by
\begin{equation}
\hat{K} = \left\{ 
\begin{array}{cc}
K & \rm{plane \: strain} \\
\frac{1-4\nu^2}{1-\nu^2} K  & \rm{plane \: stress}
\end{array} \right.
\end{equation}
Substituting the general solution for the displacement (\ref{eq:genral shp-1}) in (\ref{eq:strees cy}) enables to generally write the stress as
\begin{equation}
\sigma_{rr} = 2\mu \left(\frac{C}{3} - \frac{D}{r^2} \right) + 2 \hat{K}  C . \label{eq:stress_cyl_CD}
\end{equation}
From this we see that the solution $u = D / r$ contains only shear. However, as opposed to the spherical case discussed above, in cylindrical geometry, the solution $u = C r$ contains both shear and compression. 

Similarly to the procedure employed above for a spherical cell, we will now solve for the displacement field $u(r)$, from that we will obtain the relation between the displacement $u_0$ on the cell boundary and the active stress $\sigma_0$ that the cell applies, and from that we will calculate the elastic energy for a bounded and for an unbounded cell, and subsequently the interaction energy for displacement regulation and for stress regulation.

Inside the cell, the displacement has the same form as for the spherical geometry,
\begin{equation}
u_c = \frac{u_0 r}{R_0} . \label{eq:uc_cyl}
\end{equation}
Here, by Eq.~(\ref{eq:stress_cyl_CD}), the stress inside the cell is given by
\begin{equation}
\sigma_c = \left(\frac{2}{3}\mu_c + 2\hat{K}_c\right)\frac{u_0}{R_0} . \label{eq:sigma_c_cyl}
\end{equation}
Outside the cell, for the unbounded cylindrical geometry, the displacement is
\begin{equation}
u_{m,\infty} = \frac{u_0 R_0}{r} , \label{eq:uminf_cyl}
\end{equation}
which leads to the following radial stress on the surface of the cell ($r=R_0$)
\begin{equation}
\sigma_{m,\infty} = -2 \mu_m \frac{u_0}{R_0} , \label{eq:sigma_m_inf_cyl}
\end{equation}
and subsequently to the energy per unit length
\begin{equation}
E_\infty = 2 \pi u_0^2 \left( \frac{\mu_c}{3} + \hat{K}_c + \mu_m \right) \label{eq:E_inf_cyl} .
\end{equation}
It is interesting to note that $E_\infty$ does not depend on the cell radius, $R_0$. For the bounded geometry, the displacement is given by
\begin{equation}
u_{m,1} = \frac{u_0 R_0}{R_1^2-R_0^2} \left( \frac{R_1^2}{r} - r \right)\label{eq:um1_cyl} ,
\end{equation}
thus the stress on the surface of the cell reads
\begin{equation}
\sigma_{m,1} = - \frac{2u_0R_0}{R_1^2-R_0^2} \left[ \mu_m\left(\frac{1}{3}+\frac{R_1^2}{R_0^2}\right) + \hat{K}_m \right]  \label{eq:sigma_m_1_cyl} ,
\end{equation}
and the energy per unit length is
\begin{align}
E_1 = & \pi u_0^2 \left\{ \frac{2}{3}\mu_c + 2\hat{K}_c + \right. \nonumber\\ & \left. \left[ \frac{2}{3}\mu_m\left(1+3\frac{R_1^2}{R_0^2}\right) + 2\hat{K}_m \right] \frac{R_0^2}{R_1^2-R_0^2} \right\} . \label{eq:E_1_cyl}
\end{align}

As for the spherical case, for fixed displacement, we obtain the interaction energy by subtracting (\ref{eq:E_inf_cyl}) from (\ref{eq:E_1_cyl}), leading to 
\begin{equation}
\Delta E = \frac{2}{3}\pi u_0^2R_0^2\frac{4\mu_m+3\hat{K}_m}{R_1^2-R_0^2} . \label{eq:dE_cyl_u0}
\end{equation}
Similarly to the procedure described above for spherical symmetry, also in cylindrical symmetry for regulation of the active stress, we use (\ref{eq:sigma_c_cyl}), (\ref{eq:sigma_m_inf_cyl}) and (\ref{eq:sigma_m_1_cyl}) to obtain $u_0$ in terms of $\sigma_0$ for the bounded and for the unbounded cases, and from that obtain the interaction energy,
\begin{widetext}
\begin{equation}
\Delta E = -\frac{\pi}{6}\sigma_0^2R_0^4 \frac{\left(4\mu_m+3\hat{K}_m\right)}{\left(\mu_m+\frac{\mu_c}{3}+\hat{K}_c\right)\left[\left(\frac{\mu_m}{3}+\hat{K}_m\right)R_0^2+\mu_m R_1^2+\left(\hat{K}_c+\frac{\mu_c}{3}\right)\left(R_1^2-R_0^2\right)\right]} \label{eq:dE_cyl_s0} .
\end{equation}
\end{widetext}
As for the spherical geometry, also for the cylindrical geometry we get for displacement regulation a repulsive interaction~(\ref{eq:dE_cyl_u0}) which depends only on the elastic moduli of the matrix, while for stress regulation we obtain attractive interaction~(\ref{eq:dE_cyl_s0}), which depends also on the elastic moduli of the cell. Here there is both shear and compression in the cell, thus~(\ref{eq:dE_cyl_s0}) depends on both $\mu_c$ and $K_c$. For large separations between cells, $R_1 \gg R_0$, the interaction energy in the cylindrical geometry decays as $1/R_1^2$, as opposed to $1/R_1^3$ in spherical geometry, which is consistent with the dimensionality reduction between the two cases.

\section{Nonlinear Medium} \label{Sec:nonlinear}

\subsection{Nonlinear Material Model}

The analysis presented in Sec.~\ref{Sec:linear} assumes linear elasticity and is hence valid only for extremely small deformations. The extracellular matrix is a gel of crosslinked polymers, which exhibits nonlinear mechanical response; its response to shear is linear at small stress, while for higher stress, the differential shear modulus crosses over to a power law increase with shear stress, $G(\sigma) \propto \sigma^{3/2}$~\cite{Gardel2004, Storm2005, Vader2009, Broedersz2014}. Due to their near-incompressibility, biopolymer gels are usually compressed much less than they are sheared, thus we shall include nonlinearity only in their response to shear. Specifically, we describe the nonlinear medium surrounding our cells using the following energy density function~\cite{Shokef2012}
\begin{equation}
W=\frac{\mu}{2} \left\{ \left[1 - b \left( \bar{I}_1 - 3 \right) \right]^{-1} - 1 \right\} + \frac{K}{2} \left( J-1 \right)^2 , \label{eq:W}
\end{equation}
where $\mu$ and $K$ are the shear and bulk moduli in the linear regime, and $b$ is a dimensionless parameter quantifying the strength of the nonlinearity. The second term describes compression in a neo-Hookean manner with $J=\det(\mathbf{F})$ quantifying the compression, and $F_{i,j}=\partial x_i / \partial X_j$ the deformation gradient tensor, where $\vec{X}$ is the reference position and $\vec{x}$ the deformed position. The first term in~(\ref{eq:W}) describes shear with $\bar{I}_1= \mathrm{tr}(\mathbf{B}) / J^{2/3}$ the normalized first eigenvalue of the left Cauchy-Green strain tensor, $\mathbf{B}=\mathbf{F}\mathbf{F}^T$. Most generally, the Cauchy stress tensor is derived from the energy density as
\begin{equation}
\sigma_{ij} = \frac{1}{J} F_{ik}\frac{\partial W}{\partial F_{kj}} .
\end{equation}
However, for an energy density function which depends only on $J$ and $I_1$, this reduces to~\cite{Bower2009}
\begin{equation}
\boldsymbol{\sigma} = \frac{2}{J^{5/3}}\frac{\partial W}{\partial \bar{I}_1}\left({\bf B} - \frac{I_1}{3}{\bf 1}\right) + \frac{\partial W}{\partial J} {\bf 1} , \label{eq:sigma_gen}
\end{equation}
with ${\bf 1}$ the unit tensor. For our model (\ref{eq:W}), this yields
\begin{equation}
\boldsymbol{\sigma} = \frac{\mu}{J^{5/3}} \left[ 1 - b \left( \bar{I}_1 - 3 \right) \right]^{-2} \left({\bf B} - \frac{I_1}{3} {\bf 1}\right) + K (J-1){\bf 1} \label{eq:knowles_stress} .
\end{equation}

\begin{figure}[b]
\includegraphics[clip , trim=1cm 1.5cm 1.5cm 2cm , width=\columnwidth]{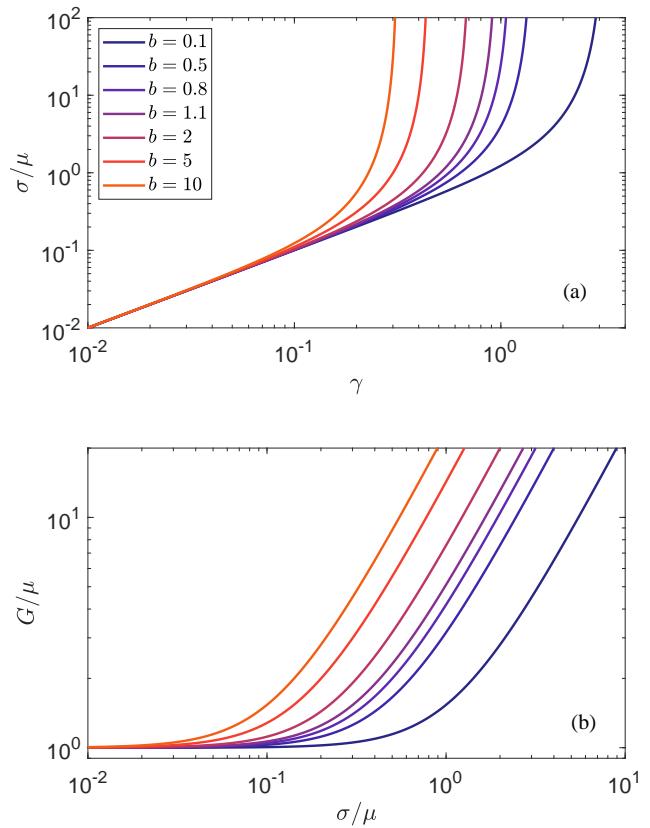}
\caption{Stress stiffenning for simple shear: a) Stress vs strain is linear for small strain and then diverges at $\gamma_a=1/\sqrt{b}$. b) differential shear modulus is constant for small stress and then increases as $G \propto \sigma^{3/2}$ for stress larger than $\mu \gamma_a$.}
\label{fig:sigmagamma}
\end{figure}

For simple shear, the deformation may be written as  $x=X+\gamma Z$, $y=Y$, $z=Z$, thus Eq.~(\ref{eq:knowles_stress}) gives a shear stress of $\sigma_{xz} = \mu \gamma \left(1-b\gamma^2\right)^{-2}$, resulting in the differential shear modulus $G \equiv d\sigma / d\gamma = \mu \left( 1 + 3b\gamma^2 \right) \left( 1 - b\gamma^2 \right)^{-3}$. In the limit of small shear $\left( \gamma \ll 1 \right)$, we obtain linear response $\sigma \approx \mu \gamma$ with a constant differential shear modulus equal to $\mu$. For large shear, the shear stress $\sigma_{xz}$ diverges as the strain $\gamma$ approaches a maximal possible strain $\gamma_a=1/\sqrt{b}$. In that limit of $\sigma \gg \mu \gamma_a$, the differential shear modulus may be approximated as $G \approx 4 \mu \left[ \sigma / \left( \mu \gamma_a \right) \right]^{3/2}$, which exhibits the power law stiffening $G \propto \sigma^{3/2}$ characteristic of biopolymer gels, see Fig.~\ref{fig:sigmagamma}.

Below, we will study the effects of this power-law stiffening on matrix mediated elastic interactions between cells. We will consider the same geometrical model studied above, of a symmetric contracting cell, surrounded by an effective wall, which provides a mean-field description for the effect of neighboring cells. For simplicity, we will restrict ourselves to a spherical cell. However, the extensions described above for for cylindrical cells are straightforward.

In order to simplify the analysis, we will restrict ourselves to the small deformations limit, which is relevant for strongly nonlinear materials ($b \gg 1$) that have a very small maximal shear strain $\gamma_a \ll 1$. In such cases, the deformations will always be small $\gamma \ll 1$, and we will have strong nonlinear effects even at small deformations. In our spherical case, this implies $u/r , du/dr \ll 1$. We will obviously keep terms linear in $u$. However, since for simple shear $b=1/\gamma_a^2$, for consistency, we will keep also terms proportional to $bu^2$, and will neglect only terms that are higher order than that. 

For spherical symmetry, the deformation is described by the difference $u(R) = r(R)-R$ between the deformed radius $r$ and the reference radius $R$. The compression and shear components of the deformation are generally given by~\cite{Bower2009}
\begin{eqnarray}
J &=& \left(1+\frac{u}{R}\right)^2\left(1+\frac{du}{dR}\right) , \nonumber \\
\bar{I}_1 &=& \left(1+\frac{u}{R}\right)^{-4/3}\left(1+\frac{du}{dR}\right)^{4/3} \nonumber \\ &+& 2\left(1+\frac{u}{R}\right)^{2/3}\left(1+\frac{du}{dR}\right)^{-2/3} . \label{eq:JI1_spher}
\end{eqnarray}
For small displacements ($u \ll R$), these are approximated as
\begin{eqnarray}
J &\approx& 1 + 2 \frac{u}{r} + \frac{du}{dr} , \nonumber \\
\bar{I}_1 &-& 3 \approx \frac{4}{3} \left( \frac{du}{dr} - \frac{u}{r} \right)^2 ,
\end{eqnarray}
where we have also used the small deformation approximation to replace $R$ with $r$.

Force balance $\nabla \cdot \boldsymbol{\sigma} = 0$ implies
\begin{equation}
\frac{d \sigma_{rr}}{d r} + \frac{2}{r} \left(\sigma_{rr} - \sigma_{\theta\theta}\right)=0 . \label{eq:equil_spher}
\end{equation}
In spherical coordinates, the general expression for the stress tensor (\ref{eq:sigma_gen}) gives 
\begin{eqnarray}
\sigma_{rr} = \frac{4}{3J^{5/3}} \frac{\partial W}{\partial\bar{I}_1} \left[ \left(\frac{du}{dr}\right)^2 + 2\frac{du}{dr} -\left(\frac{u}{r}\right)^2 - 2\frac{u}{r} \right] + \frac{\partial W}{\partial J} , \nonumber \\
\sigma_{rr} - \sigma_{\theta\theta} = \frac{2}{J^{5/3}} \frac{\partial W}{\partial\bar{I}_1} \left[ \left(\frac{du}{dr}\right)^2 + 2\frac{du}{dr} - \left(\frac{u}{R}\right)^2 - 2\frac{u}{r}\right] , \nonumber \\ \label{eq:sigmas_sp}
\end{eqnarray}
where for our strain energy density~(\ref{eq:W}),
\begin{equation}
\frac{\partial W}{\partial\bar{I}_1} = \frac{\mu}{2}\left[1-b(\bar{I}_1-3)\right]^{-2} \label{eq:dwdI1} ,
\end{equation}
and 
\begin{equation} 
\frac{\partial W}{\partial J} = K \left( \frac{du}{dr} + \frac{2u}{r} \right) . \label{eq:dwdJ}
\end{equation}
By substituting the partial differentials of the energy density function (\ref{eq:dwdI1},\ref{eq:dwdJ}) in the expression (\ref{eq:sigmas_sp}) for the stress tensor, we may write the condition for mechanical equilibrium (\ref{eq:equil_spher}) as
\begin{widetext}
\begin{eqnarray}
\left\{ K+\frac{4\mu}{3}\left[1-b\left(\bar{I}_1-3\right)\right]^{-2}\right\} \left(\frac{d^2u}{dr^2}+\frac{2}{r}\frac{du}{dr}-\frac{2u}{r^2}\right)+\frac{8\mu}{3}b\left[1-b\left(\bar{I}_1-3\right)\right]^{-3}\frac{d\bar{I}_1}{dr}\left(\frac{du}{dr}-\frac{u}{r}\right) = 0 . 
\end{eqnarray}
After multiplying by $\left[1-b\left(\bar{I}_1-3\right)\right]^3$ we write
\begin{eqnarray}
\left[1-b\left(\bar{I}_1-3\right)\right]
\left\{ K \left[1-b\left(\bar{I}_1-3\right)\right]^2 + \frac{4\mu}{3} \right\} \left(\frac{d^2u}{dr^2}+\frac{2}{r}\frac{du}{dr}-\frac{2u}{r^2}\right) + \frac{8\mu}{3} b \frac{d\bar{I}_1}{dr}\left(\frac{du}{dr}-\frac{u}{r}\right) = 0 . \label{eq:nonlin_equib}
\end{eqnarray}

To bring~(\ref{eq:nonlin_equib}) to dimensionless form, we normalize the radial coordinate by the cell radius $\tilde{r}=r/R_0$ and the displacement by the displacement on the cell boundary $\tilde{u}=u/u_0$. We quantify the nonlinearity by $A = b \left( u_0 / R_0 \right)^2 = \left[\left(u_0/R_0\right)/\gamma_a\right]^2$, which is a measure of the ratio between the typical scale $u_0/R_0$ of the strain on the cell surface and the maximal shear strain $\gamma_a = 1/\sqrt{b}$ that the material can sustain. Finally, we express the ratio of bulk modulus to linear shear modulus in terms of the Poisson ratio~(\ref{eq:Poisson}), and eventually write the equilibrium condition as:
\begin{eqnarray}
 \left(1-A\tilde{I}_3\right) \left[ \frac{1+\nu}{2(1-2\nu)} \left(1-A\tilde{I}_3\right)^2+1 \right] \left(\frac{d^{2}\tilde{u}}{d\tilde{r}^{2}}+\frac{2}{\tilde{r}}\frac{d\tilde{u}}{d\tilde{r}}-2\frac{\tilde{u}}{\tilde{r}^{2}}\right) + 2A\frac{d\tilde{I_3}}{d\tilde{r}}\left(\frac{d\tilde{u}}{d\tilde{r}}-\frac{\tilde{u}}{\tilde{r}}\right)=0 , \label{eq:nonlin_equib_dimesionless}
\end{eqnarray}
\end{widetext}
where 
\begin{eqnarray}
\tilde{I}_{3} = \left(\frac{R_0}{u_0}\right)^2 \left(\bar{I}_1-3\right) = \frac{4}{3}\left(\frac{d\tilde{u}}{d\tilde{r}}-\frac{\tilde{u}}{\tilde{r}}\right)^2  \label{eq:Itilde3}
\end{eqnarray}
is a dimensionless measure of the shear in the matrix.

Equation~(\ref{eq:nonlin_equib_dimesionless}) is a nonlinear second order ordinary differential equation for the dimensionless displacement as a function of the dimensionless radius, $\tilde{u}\left(\tilde{r}\right)$. We obtain $\tilde{u}\left(\tilde{r}\right)$ by numerically solving the boundary value problem using our two boundary conditions: given displacement on the cell border $\tilde{u}(1)=1$, and zero displacement, either at infinity for the unbounded geometry $\tilde{u}(\infty)=0$ or at $\tilde{R}_1=R_1/R_0$ for the bounded geometry $\tilde{u}\left(\tilde{R}_1\right)=0$.

Clearly, in the limit $A=0$ of vanishing nonlinearity, Eq.~(\ref{eq:nonlin_equib_dimesionless}) reduces to Eq.~(\ref{eq:ode sph}), which describes a linearly elastic medium. In this linear limit, there are two solutions, $\tilde{u}=\tilde{r}$ and $\tilde{u}=1/\tilde{r}^2$, and as shown in Sec.~\ref{Sec:linear}, a linear combination of the two can match any given set of boundary conditions. More interestingly, in this spherical geometry, the solution $\tilde{u}=\tilde{r}$ that contains only compression without shear, leads to $\tilde{I}_3=0$  and thus solves Eq.~(\ref{eq:nonlin_equib_dimesionless}) for arbitrary strength $A$ of the nonlinearity. This solution is consistent with the first boundary condition at the cell border $\tilde{u}(1)=1$, but it increases in magnitude with $\tilde{r}$ and thus cannot satisfy the second, zero displacement boundary condition at $\tilde{r}=\infty$ or at $\tilde{r}=\tilde{R}_1$.

\subsection{Unbounded Cell}

\begin{figure}[t]
\includegraphics[clip , trim=0.2cm 0.5cm 1cm 1cm , width=\columnwidth]{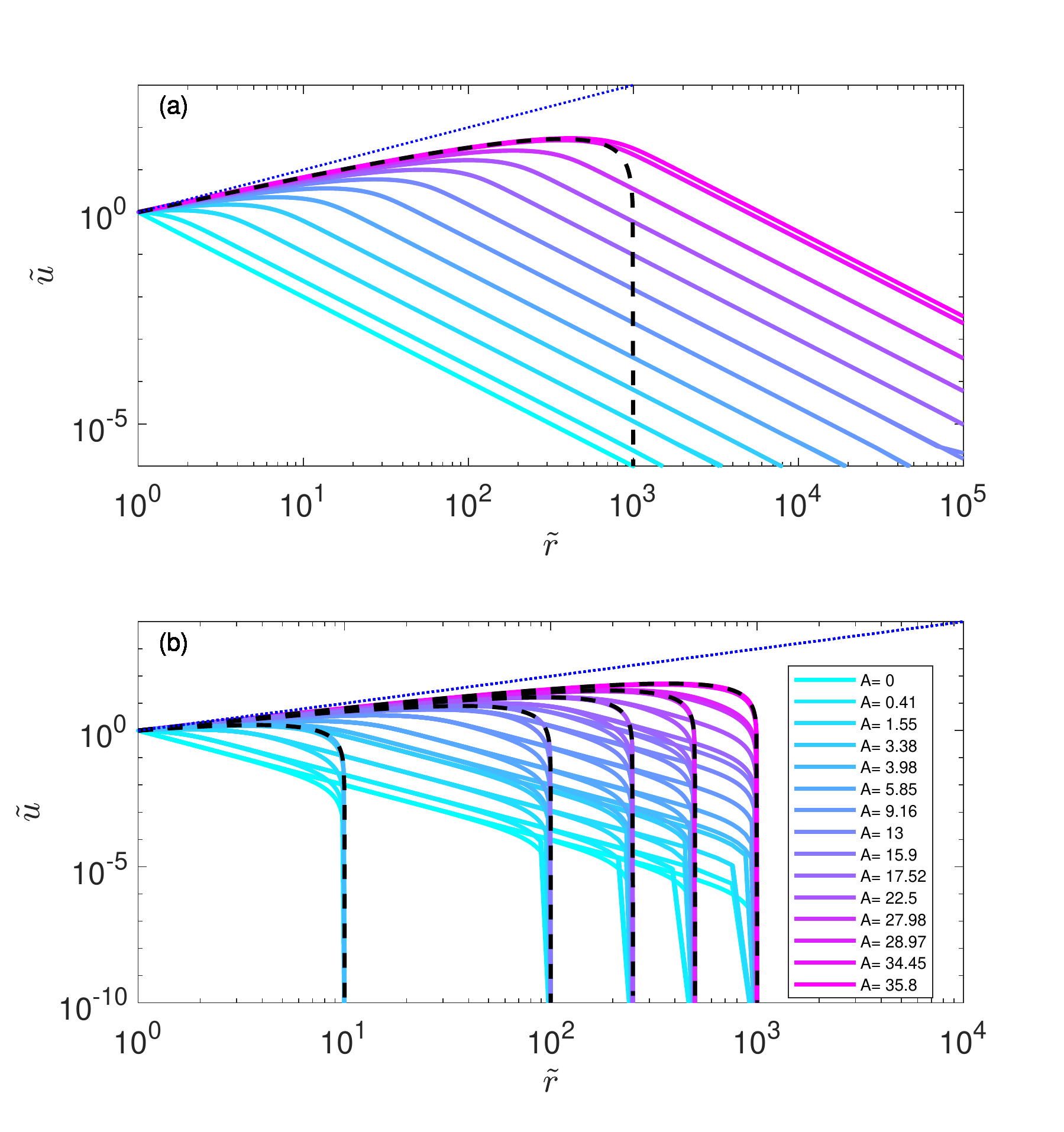}
\caption{Normalized displacement $\tilde{u}=u/u_0$ vs. normalized radius $\tilde{r}=r/R_0$ in spherical nonlinear medium for the (a) unbounded geometry and (b) geometry bounded at $\tilde{R}_1=R_1/R_0=10,100,250,500,1000$. Colors mark the dimensionless strength $A=bu_0^2/R_0^2$ of the nonlinearity, as indicated in the legend. All results are for Poisson ratio $\nu=0.4$. In both geometries, for increasing nonlinearity, the near-field behavior tends toward the shear-less solution $\tilde{u}=\tilde{r}$ represented by the blue dotted line, and may be approximated by Eq.~(\ref{eq:u_approx}), which is drawn in a black dashed line only for the maximal $A$ value for each case.}
\label{fig:u_vs_r}
\end{figure}

For an unbounded cell, the dimensionless displacement $\tilde{u}\left(\tilde{r}\right)$ depends only on the dimensionless strength $A$ of the nonlinearity and on the Poisson ratio $\nu$. The behavior of an unbounded cell in this nonlinear material model depends very weakly on the Poisson ratio~\cite{Shokef2012}, thus we will not consider this dependence here, and all the numerical results we present below are for $\nu=0.4$. As shown in Fig.~\ref{fig:u_vs_r}a, for $A=0$ the response is completely linear, and the displacement decays according to the prediction of linear elasticity, $\tilde{u}\left(\tilde{r}\right)=1/\tilde{r}^2$, or $u(r)=u_0 \left(R_0/r\right)^2$. As $A$ increases, an increasingly larger region near the cell responds nonlinearly, approaching the shear-less solution $\tilde{u}\left(\tilde{r}\right)=\tilde{r}$, or $u(r)=u_0 r/R_0$, while far enough from the cell $\tilde{r} \gg 1$, deformations decay and the matrix restores its linear response, leading to a $\tilde{u} \propto 1/\tilde{r}^2$ decay of displacements in the far field. The far-field behavior may be written as $u(r)=u_{\rm eff} \left(R_0/r\right)^2$, or $\tilde{u}\left(\tilde{r}\right)=\tilde{u}_{\rm eff}/\tilde{r}^2$, where $\tilde{u}_{\rm eff}=u_{\rm eff}/u_0$ describes the amplification of the effective displacement on the cell border as felt at long distances. We will denote by $\tilde{R}_* \equiv R_*/R_0$ the position of the maximum of $\tilde{u}(\tilde{r})$. This position roughly divides space into the nonlinear near-field compression-dominated regime at $\tilde{r}<\tilde{R}_*$ and the linearly-elastic far-field shear-dominated regime at $\tilde{r}>\tilde{R}_*$. 

\subsection{Bounded Cell}

To study matrix-mediated mechanical interaction between cells in a nonlinear medium, we employ the mean-field geometrical setup presented in Sec.~\ref{Sec:Model}, which we have studied for a linear matrix in Sec.~\ref{Sec:linear}. We now numerically solve Eq.~(\ref{eq:nonlin_equib_dimesionless}) with the boundary condition $\tilde{u}(\tilde{R}_1)=0$. In this bounded geometry, the dimensionless displacement solution $\tilde{u}\left(\tilde{r}\right)$ depends not only on the Poisson ratio $\nu$ and on the dimensionless nonlinearity $A$, but also on the dimensionless distance $\tilde{R}_1$ between cells. We show in Fig.~\ref{fig:u_vs_r}b numerical results for varying $A$ and $\tilde{R}_1$. 

The first thing to note in comparison to a linear medium is the following: In the unbounded geometry, the dimensionless nonlinearity $A$ of the problem may take arbitrarily large values. This means that for a given material nonlinearity $b$, the displacement $u_0$ on the cell boundary can take any value. As seen in Fig.~\ref{fig:u_vs_r}a, increasing $A$ causes an increase in the radius $\tilde{R}_*$ where $\tilde{u}\left(\tilde{r}\right)$ crosses over from the compression-only solution $\tilde{u}=\tilde{r}$ to the shear-only solution $\tilde{u}=\tilde{u}_{\rm eff}/\tilde{r}^2$. However, for the bounded geometry, this crossover length $\tilde{R}_*$ clearly cannot be larger than $\tilde{R}_1$. This implies that for given dimensionless cell-cell distance $\tilde{R}_1$, there is a maximal value $A_a$ that the nonlinearity can reach. Thus for given geometry specified by $R_0$ and $R_1$ and for given material nonlinearity $b$, the displacement $u_0$ on the cell boundary can increase only up to some maximal value $u_a = R_0 \sqrt{A_a ( \tilde{R}_1 ) /b}$.

In our nonlinear model~(\ref{eq:W}), for large strain the resistance to shear becomes increasingly larger than the resistance to compression. Hence, in mechanical equilibrium, the shear component of the deformation tends to be minimal. For our spherical geometry, requiring $\tilde{I}_3=0$ leads to the solution $\tilde{u}\left(\tilde{r}\right)=\tilde{r}$, which is consistent with the boundary condition $\tilde{u}(1)=1$, but cannot be consistent with the boundary condition $\tilde{u}\left(\infty\right)=0$ or $\tilde{u}(\tilde{R}_1)=0$. An approximate solution to~(\ref{eq:nonlin_equib_dimesionless}) that decays in the far field is obtained by taking $\tilde{I}_3=1/A$~\cite{Shokef2012}, which by substitution in~(\ref{eq:Itilde3}) leads to 
\begin{equation}
\tilde{u}\left(\tilde{r}\right) = \tilde{r} - \sqrt{\frac{3}{4A}} \tilde{r} \log{\tilde{r}} , \label{eq:u_approx}
\end{equation}
which in turn also causes $d\tilde{I}_3/d\tilde{r}$ to vanish. Figure~\ref{fig:u_vs_r} shows how well~(\ref{eq:u_approx}) agrees with the near-field behavior of the displacement field, both for the bounded and for the unbounded geometry.

\begin{figure}[t]
\includegraphics[clip , trim=0.1cm 1.3cm 0.7cm 1.3cm , width=\columnwidth]{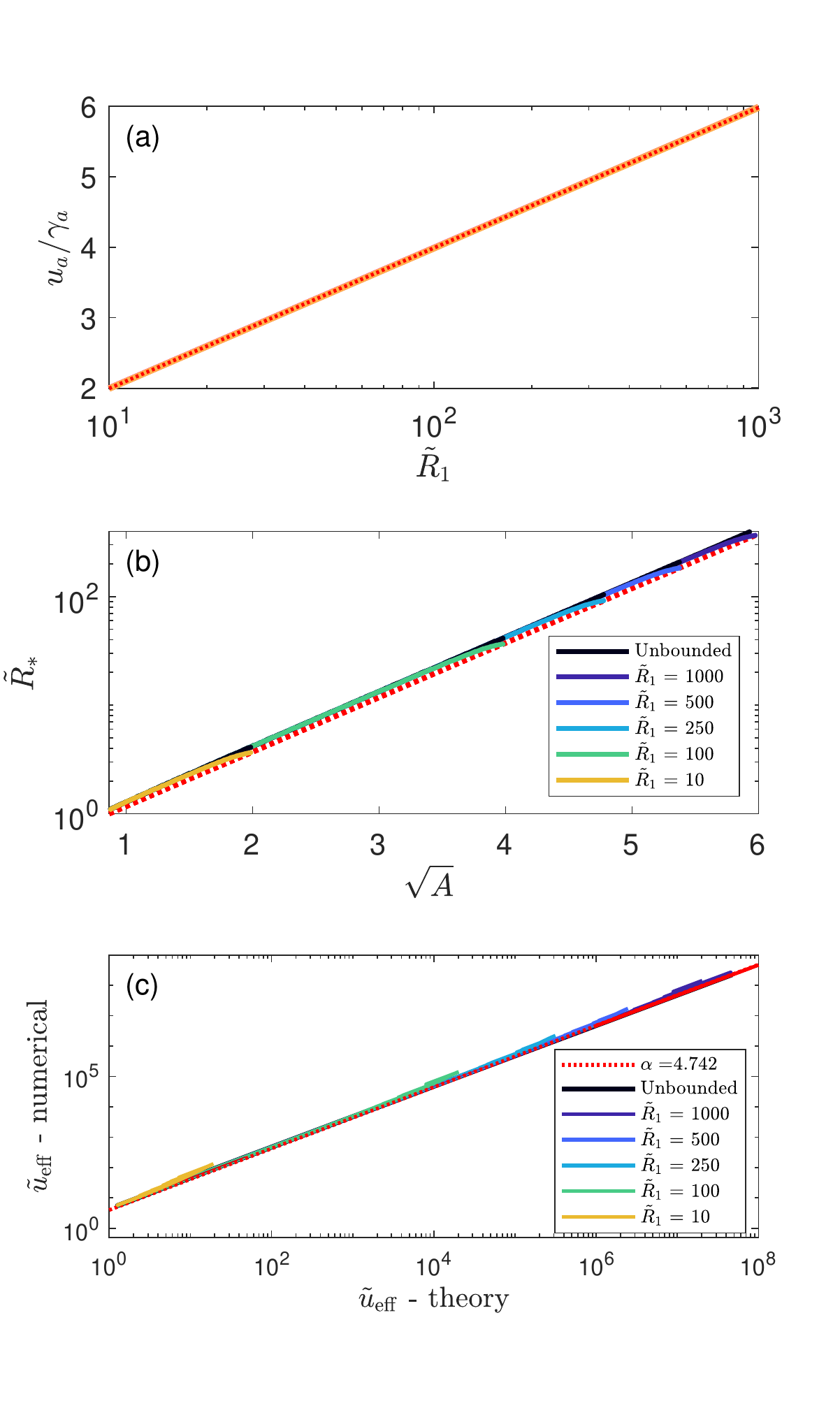}
\caption{(a) Maximal possible cell contraction increases with cell-cell distance according to Eq.~(\ref{eq:u0max}). Numerical results for $b=0.1,0.5,0.8,1.1,2,5,10$ all collapse to a single curve, which perfectly agrees with the theoretical prediction. (b) Position of maximal displacement increases with dimensionless nonlinearity according to Eq.~(\ref{eq:Rstar}) both for the bounded and for the unbounded geometries. (c) Effective far field displacement increases with $A$ and decreases with $\tilde{R}_1$ following Eq.~(\ref{eq:ueff_vs_R1}). In all panels, theoretical prediction is marked by red dotted line.}
\label{fig:Rstar_and_u0max}
\end{figure}

We may approximate the maximal possible displacement $u_a$ on the cell boundary by requiring that for $A_a$ the approximate near-field solution~(\ref{eq:u_approx}) reaches zero at the distance $\tilde{R}_1$. The justification for this is that for $u_0<u_a$, or correspondingly $A<A_a$, the displacement field has room to cross over at larger distances to the linear elasticity solution so that it will eventually satisfy the boundary condition $\tilde{u}(\tilde{R}_1)=0$. For $u_0>u_a$ on the other hand, the near-field solution~(\ref{eq:u_approx}) is still finite at $\tilde{R}_1$ and the zero-displacement boundary condition there may not be satisfied. Thus, setting $\tilde{u}\left(\tilde{R}_1\right)=0$ in~(\ref{eq:u_approx}), we write 
\begin{equation}
\tilde{R}_1 - \sqrt{\frac{3}{4A_a}} \tilde{R}_1 \log{\tilde{R}_1} = 0 ,
\end{equation}
from which we obtain the maximal dimensionless nonlinearity
\begin{equation}
A_a = \frac{3}{4} \log^2 \tilde{R_1}.
\end{equation}
This lead to the maximal possible cell contraction
\begin{equation}
\tilde{u}_a = \frac{u_a}{R_0} = \sqrt{\frac{3}{4}} \gamma_a \log{\frac{R_1}{R_0}} , \label{eq:u0max}
\end{equation}
which scales linearly with the maximal shear strain $\gamma_a$, and logarithmically with the dimensionless cell-cell distance. This simple theoretical prediction agrees remarkably well with our numerical results, as shown in Fig.~\ref{fig:Rstar_and_u0max}a.  Note that at $A=A_a$, our approximate solution~(\ref{eq:u_approx}) becomes exact, as it satisfies the boundary conditions both at $\tilde{r}=1$ and at $\tilde{r}=\tilde{R}_1$, see Fig.~\ref{fig:u_vs_r}b. Before moving on to discussing the interaction energy we also note that~(\ref{eq:u0max}) may be inverted to obtain the closest distance between cells that contract by a given amount $u_0$,
\begin{equation}
\min(R_1) = R_0 \exp{\left(\sqrt{\frac{4}{3}} \frac{u_0/R_0}{\gamma_a} \right)} ,
\end{equation}
or alternatively
\begin{equation}
\min\left(\tilde{R}_1\right) = \exp{\left(\sqrt{\frac{4A}{3}} \right)} .
\end{equation}

\subsection{Interaction Energy}

To calculate the interaction energy for given cell size $R_0$, cell-cell distance $R_1$, material nonlinearity $b$, and cell contraction $u_0$, we need to first numerically solve the boundary value problem to get the displacement field $u(r)$ for the bounded and for the unbounded geometries for any intermediate cell contraction $0<w<u_0$. Then, we use $u$ and $du/dr$ at $R_0$ to evaluate the radial stress~(\ref{eq:sigmas_sp}) $\tau(w)=\sigma_{rr}(R_0)$ on the cell surface. Finally, we numerically integrate Eq.~(\ref{Eq:work-def}) to obtain the elastic energy stored in the medium for the bounded and for the unbounded cases, and subtract them to get the interaction energy.

\begin{figure}[b]
\includegraphics[clip , trim=0.5cm 1cm 1.2cm 1.1cm , width=\columnwidth]{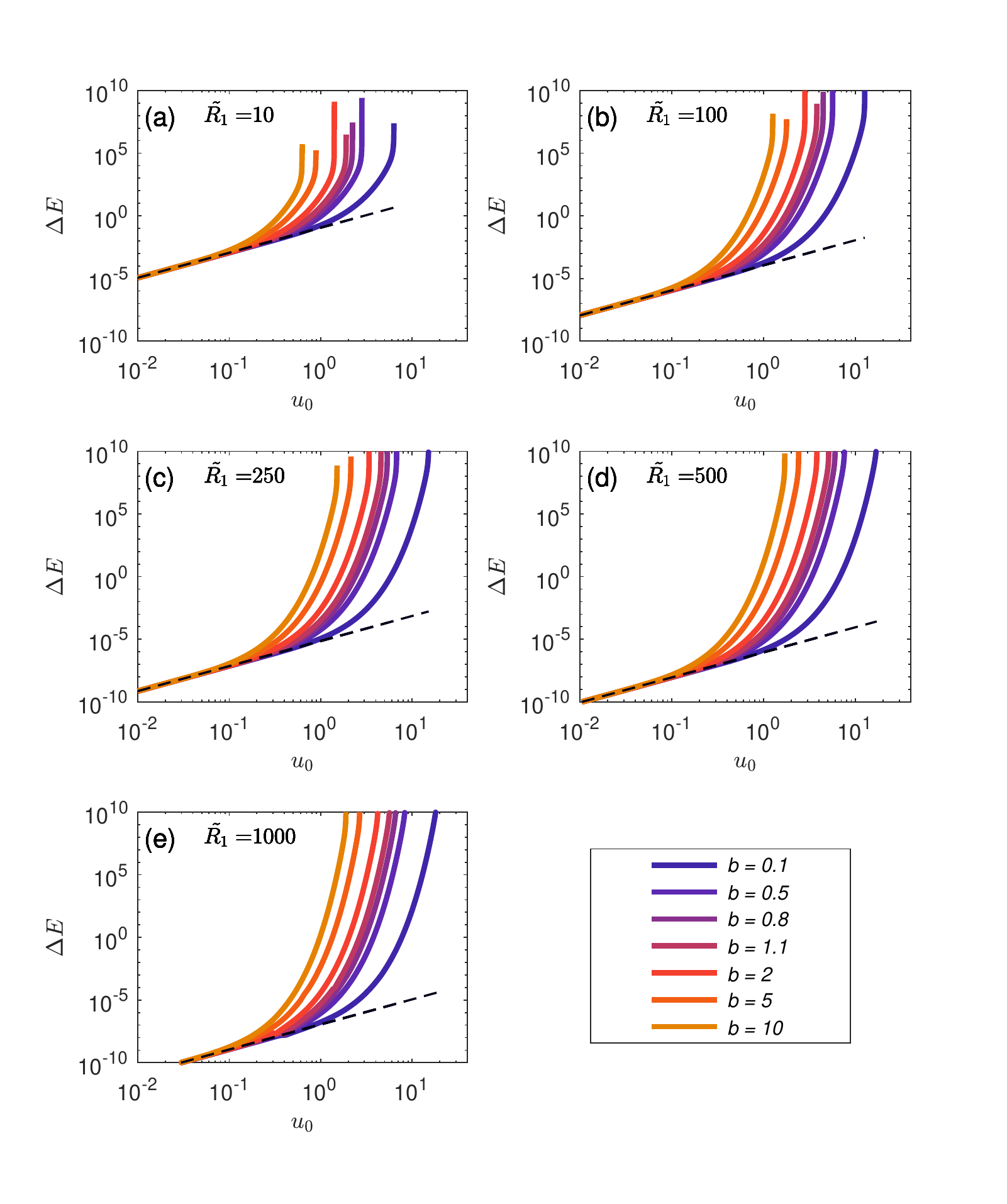}
\caption{Interaction energy vs. cell contraction for different values of $R_1$ and $b$. Interaction energy is scaled by $\mu R_0$, and cell displacement is scaled by $R_0$. Black dashed lines are linear elasticity behavior for $b=0$.}
\label{fig:energy}
\end{figure}

\subsubsection{Displacement Regulation}

\begin{figure}[b]
\includegraphics[clip , trim=0cm 0.1cm 1.1cm 0.6cm , width=\columnwidth]{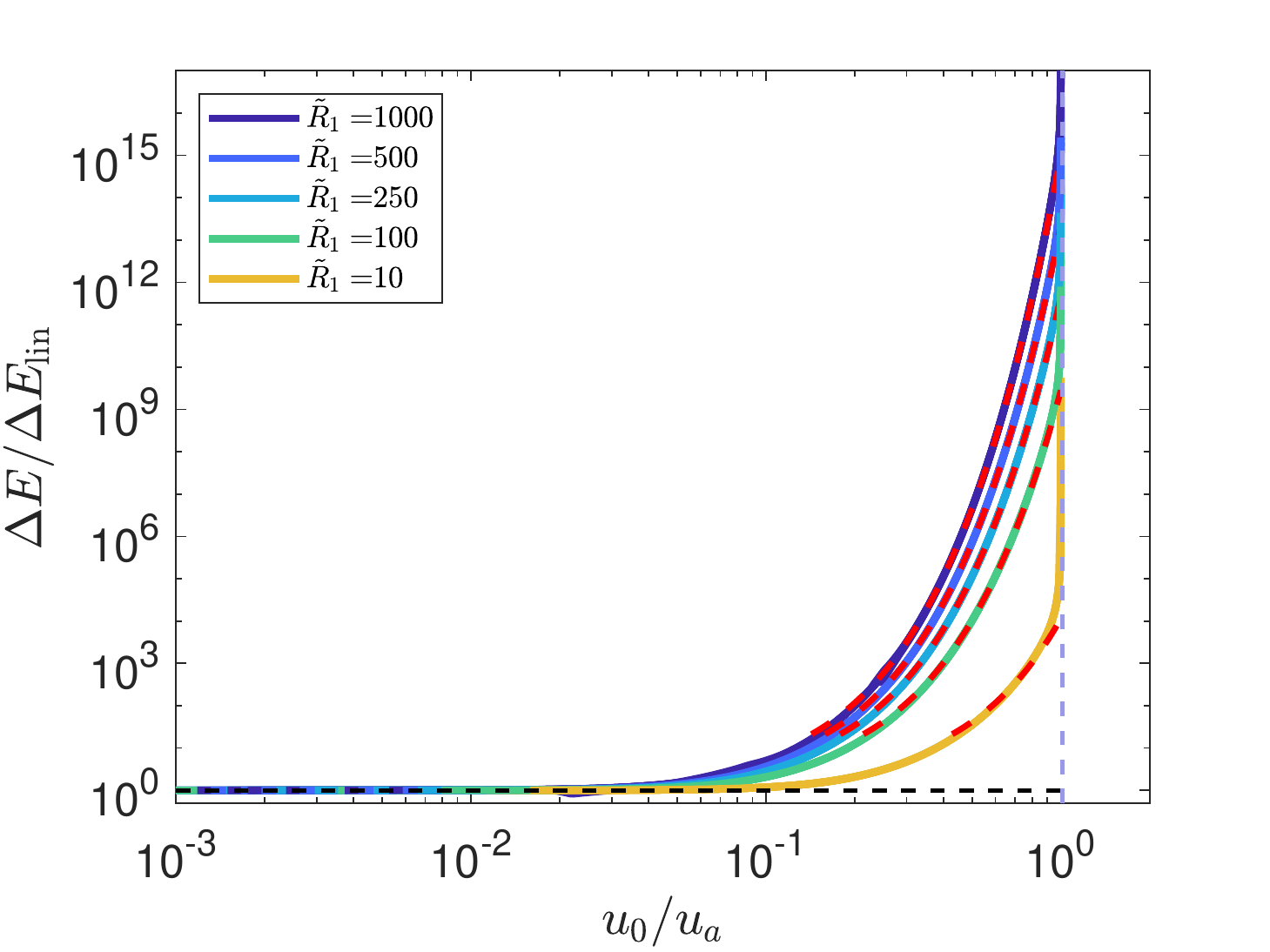}
\caption{Scaled interaction energy behaves according to linear elasticity~(\ref{eq:eee}) for small cell contraction (black dashed line). As $u_0$ approaches $u_a$, the interaction energy grows according to Eq.~(\ref{eq:nonlin_dE}) (red dashed lines), until it diverges at $u_0=u_a$ (blue dashed line). Results for different values of the material nonlinearity $b$ collapse to curves that depend only on $\tilde{R}_1$.}
\label{fig:energy_scaled}
\end{figure}

The numerical results of $\Delta E$ vs $u_0$ for various values of $b$ and $R_1$ are given in Fig.~\ref{fig:energy}. For small $u_0$, the material responds linearly, thus the interaction energy grows quadratically with $u_0$, as predicted in a linear medium. For increasing $u_0$, the interaction energy grows more rapidly and diverges at a finite contraction $u_a$, which depends on $R_1$ and on $b$ according to~(\ref{eq:u0max}). We first note that the dependence of $\Delta E$ on the material nonlinearity $b$ may be scaled out if we normalize $\Delta E$ by the interaction energy~(\ref{eq:eee}) that we would get for these values of $u_0$, $R_0$ and $R_1$ in a linear medium. Figure~\ref{fig:energy_scaled} shows that after this normalization, and by also scaling $u_0$ by its maximal possible value $u_a$ in the nonlinear case, we obtain collapse of the results for different $b$ values to master curves that depend only on $\tilde{R}_1 = R_1 / R_0$.

\begin{figure}[b]
\includegraphics[clip , trim=0.2cm 0.2cm 1cm 0.5cm , width=\columnwidth]{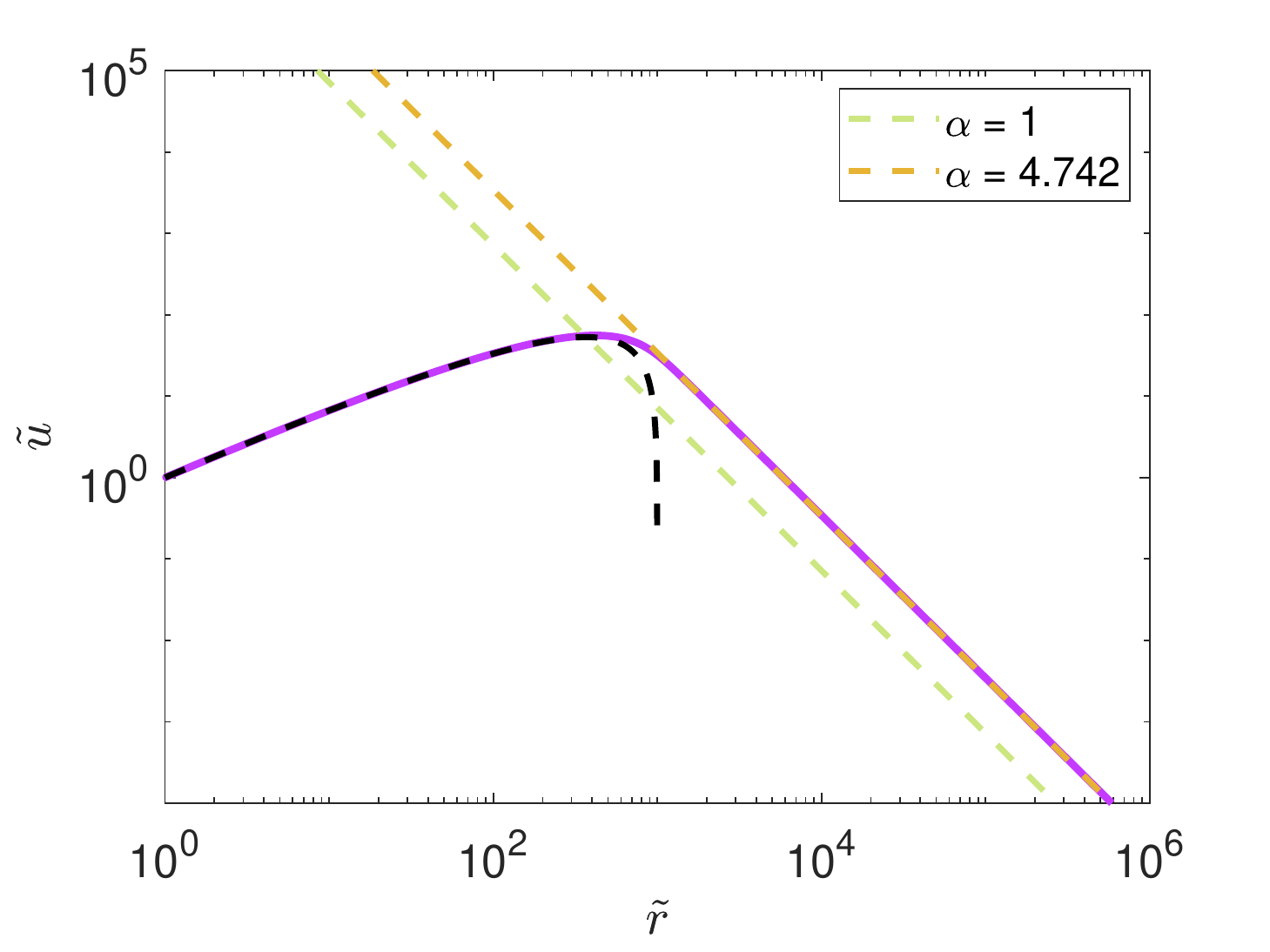}
\caption{Crossover of $\tilde{u}(\tilde{r})$ from near-field behavior of Eq.~(\ref{eq:u_approx}) (dashed black line) to far field $\tilde{u}=\tilde{u}_{\rm eff}/\tilde{r}^2$ behavior. Matching the two solutions at $\tilde{R}_*$ (green dashed line) underestimates $\tilde{u}_{\rm eff}$. Matching at a slightly larger distance (orange dashed line) is obtained by taking $\alpha>1$ in Eq.~(\ref{eq:ueff_vs_R1}). Results are shown for A=35.7858.}
\label{fig:crossover}
\end{figure}

To approximate the entire dependence on $u_0$ as well as the dependence on $\tilde{R}_1$, we first note that the near-field approximate solution~(\ref{eq:u_approx}) is maximal at 
\begin{equation}
\tilde{R}_* = \exp{\left(\sqrt{\frac{4A}{3}} - 1 \right)} . \label{eq:Rstar}
\end{equation}
This agrees very well with the position of the maximum of $\tilde{u}(\tilde{r})$ as obtained from our numerical solution, both for bounded and for unbounded geometries, see Fig.~\ref{fig:Rstar_and_u0max}b. Note that (\ref{eq:u_approx}) has a local maximum at $\tilde{r}>1$ only for $A> 3/4$, thus all the analysis that follows is relevant only for dimensionless nonlinearity that is larger than this level. Now, in the far field, we expect the solution~(\ref{eq:SSE}) in a linear medium to hold, but with the displacement $u_0$ on the surface of the cell replaced by some larger, effective displacement $u_{\rm eff}$. In dimensionless form this read
\begin{equation}
\tilde{u}\left(\tilde{r}\right) = \frac{\tilde{u}_{\rm eff}}{\tilde{R}_1^3-1} \left( \frac{\tilde{R}_1^3}{\tilde{r}^2} - \tilde{r} \right) \label{eq:ffu} .
\end{equation}
As has successfully been done for the unbounded geometry~\cite{Shokef2012}, also for our bounded geometry, the dimensionless effective displacement may be obtained by matching the near field solution~(\ref{eq:u_approx}) and the far field solution~(\ref{eq:ffu}) at $\tilde{R}_*$~(\ref{eq:Rstar}). The crossover between these two solutions is not sharp but exists over a certain crossover region, see Fig.~\ref{fig:crossover}. Thus we allow for a dimensionless multiplicative factor $\alpha$ between the two solutions at $\tilde{R}_*$. This leads to
\begin{equation}
\tilde{u}_{\rm eff} = \frac{u_{\rm eff}}{u_0} = \alpha \sqrt{\frac{3}{4A}} \frac{\tilde{R}_1^3-1}{\frac{\tilde{R}_1^3}{\tilde{R}_*^3}-1} . \label{eq:ueff_vs_R1}
\end{equation}
For $\tilde{R}_1 \gg \tilde{R}_*$ this reduces to $\tilde{u}_{\rm eff}=\alpha\sqrt{3/(4A)}\tilde{R}_*^3$. Note that~(\ref{eq:ueff_vs_R1}) cannot explain the divergence of $\Delta E$ as $u_0$ approaches $u_a$ since it is not singular at the maximal cell contraction $A_a$; it is just equal to 
\begin{equation}
\tilde{u}_{\rm eff} = \alpha \frac{\tilde{R}_1^3-1}{\log\tilde{R}_1 \left(e^3-1\right)} .
\end{equation}
Figure~\ref{fig:Rstar_and_u0max}c shows the agreement of~(\ref{eq:ueff_vs_R1}) with the numerical results, and identifies the value $\alpha \approx 4.7$ for all the values of $b$ and $\tilde{R}_1$ considered.
 
The near field behavior of the displacement field is very similar in the bounded and in the unbounded geometries, and most of the differences between the two cases are in the far field (see Fig.~\ref{fig:u_vs_r}). Thus we assume that most of the contribution to the difference in the elastic energy stored in the medium between the two geometries comes from the far field region. This enables us to approximate the interaction energy as the difference in energies stored in a linearly elastic medium displaced by $u_{\rm eff}$ rather than by $u_0$ at the cell boundary. Thus we substitute $u_{\rm eff}$ of the bounded geometry in Eq.~(\ref{eq:eee}) for $\Delta E$, and get
\begin{equation}
\Delta E = \Delta E_{\rm lin} \frac{3\alpha^2}{4A} \left( \frac{\tilde{R}_1^3-1}{\frac{\tilde{R}_1^3}{\tilde{R}_*^3}-1} \right)^2 , \label{eq:nonlin_dE}
\end{equation}
where $\Delta E_{\rm lin}$ is the interaction energy in a linear medium~(\ref{eq:eee}). Figure~\ref{fig:energy_scaled} shows the impressive agreement of this theoretical expression with the numerical results up until very close to the divergence of $\Delta E$ at $u_a$.

\subsubsection{Stress Regulation}

We now numerically evaluate the radial stress~(\ref{eq:sigmas_sp}) on the surface of the cell for the unbounded and for the bounded geometries, and subtract the corresponding energies at given stress to obtain the interaction energy for the situation in which the cell regulates the stress on its surface regardless of its mechanical environment. For simplicity, for cells in a nonlinear matrix we will assume that the cell does not resist the deformation and that all the active force applied by the cell is directed toward generating stress in the matrix. Namely $\sigma_c=0$ and $\sigma_0=-\sigma_m$. When relating this to the linear elasticity solution~(\ref{eq:destressridis}), this is obtained by setting $K_c=0$.

The reasoning explaining why for displacement regulation the interaction energy is positive (repulsion) and for stress regulation it is negative (attraction) is valid also for cells in a strain stiffening matrix. Our numerical calculations indeed result in positive interaction energy for displacement regulation and negative interaction energy for stress regulation. For displacement regulation we saw that the divergence of the shear stress at a finite strain $\gamma_a$ (see Fig.~\ref{fig:sigmagamma}a) causes the interaction energy to diverge at a finite displacement $u_a$. For stress regulation on the other hand, at least from the point of view of the matrix resistance, the active stress that the cell generates can be arbitrarily large (see Fig.~\ref{fig:sigmagamma}b). 

\begin{figure}[t]
\includegraphics[clip , trim=2cm 2cm 2cm 2cm , width=\columnwidth]{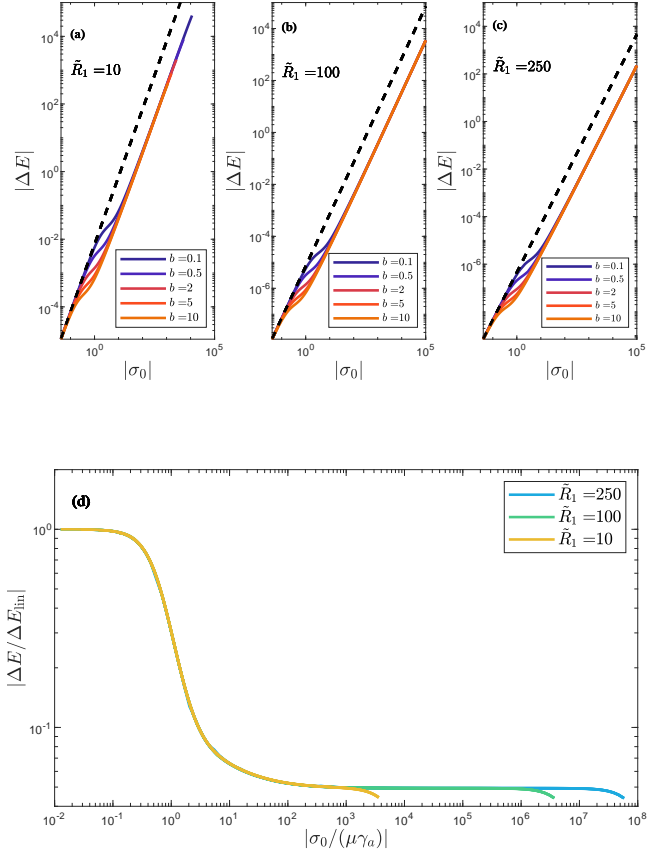}
\caption{Interaction energy with stress regulation in a nonlinear medium. (a-c) Results for different values of $\tilde{R}_1$ and $b$. Dashed black lines are linear elasticity behavior~(\ref{eq:destressridis}). Interaction energy is scaled by $\mu / R_0^3$ and stress is scaled by $\mu$. (d) When normalizing the interaction energy by the linear prediction and normalizing the stress by the typical stress $\mu \gamma_a$ where strain stiffening becomes dominant, results for all $b$ and $\tilde{R}_1$ collapse to a single curve.}
\label{fig:stress_reg}
\end{figure}

When plotting in Fig.~\ref{fig:stress_reg} the numerically obtained interaction energy vs. the regulated stress we see the linear-elasticity $\Delta E \propto - \sigma_0^2$ behavior~(\ref{eq:destressridis}) at low stress. This is followed by a more complex crossover at intermediate non-linearity, and eventually at extremely large stress the interaction energy again scales as $\sigma_0^2$, enabling arbitrarily large values of the stress, as expected. Interestingly, we obtain the quadratic scaling $\Delta E \propto \sigma_0^2$ not only in the linear regime but also in the strongly nonlinear regime. This may be explained from the fact that the nonlinear dependence of $u_{\rm eff}$ on $u_0$ and of $\sigma_0$ on $u_{\rm eff}$ are reciprocal such that $\sigma_0 \propto u_{\rm eff}$~\cite{Shokef2012}. Combining this with the result presented above that $\Delta E \propto u_{\rm eff}^2$ leads to $\Delta E \propto \sigma_0^2$. Impressively, when we scale the nonlinear interaction energy by the linear result, and scale the stress by the typical stress $\mu \gamma_a$ where the strain stiffening becomes significant, all the numerical results collapse to a single curve, as shown in Fig.~\ref{fig:stress_reg}d for multiple values of $\tilde{R}_1$ and $b$.

\section{Discussion} \label{Sec:discussion}

In this paper, we used a mean-field approach to study matrix mediated interactions between contracting biological cells in linear and in nonlinear elastic surroundings. We showed how the regulatory behavior of the cell's mechanical activity always gives rise to repulsion when displacement is regulated and to attraction when stress is regulated. For displacement regulation, the interaction does not depend on the rigidity of the cell. For stress regulation, it depends only on the bulk modulus of the cell in spherically symmetric situations, while in cylindrical setups it depends also on its shear modulus. For a nonlinear, shear stiffening matrix, the interaction energy diverges at a finite displacement, while for stress regulation the stress may be arbitrarily large.

For simplicity, we treated the cell as a passive, linearly elastic solid. This is clearly a very crude description, and the way that cells set their size, displacement and forces include many further processes, see e.g. \cite{Adar2020, Doss2020}. However, we emphasize that for displacement regulation the mechanical response of the cell is irrelevant for the interaction energy. Namely, no matter how complicated is the response of the cell, if we consider the situation in which cells regulate the displacement they generate, then the inside of the cell behaves exactly the same regardless of the distance to neighboring cells, and our results both for a linearly elastic matrix and for a nonlinear matrix are valid also beyond the simplified assumption that the cell is a passive, linearly elastic solid. In the case of stress regulation, our analytical results for a linearly elastic matrix rely on the linearity assumption for the cell, whereas for a nonlinear matrix we assumed that the cell is much softer than the matrix, and therefore its mechanical response does not affect the interaction with other cells.

We focused our nonlinear material model for the extracellular matrix on shear stiffening and included nonlinearity only in the resistance to shear. It would be interesting to study the interactions and different regulatory behaviors that we consider also in models that take into account the anisotropy of the matrix~\cite{Xu2015,Wang2020,Goren2020}, that include also nonlinearity in the compressive response, and in models that take into account the discrete fibrous nature of the biopolymer gel comprising the extracellular environment~\cite{Sopher2018,Goren2020,Natan2021,Alisafaei2021}.

\begin{acknowledgements}
We thank Dan Ben-Yaakov, Roman Golkov, Shahar Goren, Yoni Koren, Ayelet Lesman and Sam Safran for fruitful discussions. This research was supported in part by a grant from the United States-Israel Binational Science Foundation, by the Israel Science Foundation Grant No. 968/16 and by the National Science Foundation Grant No. NSF PHY-1748958.
\end{acknowledgements}

\end{document}